# Effects of metals (X = Zn, Co) on structure, electronic bands and gravimetric capacity of KXH$_3$ hydrides

Anupam[1], Shyam Lal Gupta[2], Sumit Kumar[3], Samjeet Singh Thakur[4], Sanjay Panwar[1], Diwaker[5,*]

[1]School of Basic and Applied Sciences, Maharaja Agrasen University, Baddi, Solan, 174103, H P, INDIA
[2]Department of Physics, HarishChandra Research Institute, Prayagraj, Allahabad, 211019, U P, INDIA
[3]Department of Physics, Government College, Una, 174303, H P, INDIA
[4]Department of Chemistry, NSCBM, Government College, hamirpur, Hamirpur, 177005, H P, INDIA
[5]Department of Physics, SCVB Government College, Palampur, Kangra, 176061, H P, INDIA

diwakerphysics@gmail.com

**Abstract.** Using the WIEN2K code, the hydrogen storage capabilities of lithium-based KXH$_3$ (X = Zn, Co) hydrides perovskites are examined. To verify the stability of these hydrides, first-principles simulations are employed to examine their structural, electronic, and hydrogen storage capabilities. These compositions' structural investigation shows that the hydrides are stable and part of the cubic space group (221 Pm-3m). We have examined several aspects of these composition's features throughout, using the Perdew-Burke-Ernzerhof generalized gradient approximation. The study identifies stable phases and structural parameters of hydrides using B-E equations, assessing thermodynamic stability in terms of hydrogen storage capacities. The metallic nature of these hydrides is confirmed through band structure and density calculations using WIEN2K.

## 1 Introduction

Perovskites, with their exceptional photoelectric and catalytic capabilities, have gained significant interest in energy storage, pollutant degradation and optoelectronic devices in recent decades.[1]. Perovskites are any materials whose crystal structure has the formula ABX$_3$. These materials can be further classified into two categories: inorganic perovskite and organic-inorganic hybrid perovskite. In these cases, the anion X is a member of the halogen or chalcogen group, and the two metal cations, A and B have distinct sizes. and belong to the alkali and alkaline earth metals, respectively. SrTiO$_3$, having a structure that seems to be crystalline and disordered in the lower symmetric space group, is considered to be a superb example of a perovskite. Because of their unique combination of magnetic and electric characteristics, perovskites are fascinating and important for technological applications. Perovskites hydrides represented by ABH$_3$, with a band gap of at least 2 eV, consist of alkali and alkaline earth metals at poistion A and B respectvely. Goldschmidt's idea of tolerance factors is mostly supported by their stability trend. ABH$_3$ may only be produced in the following scenarios: B must be a transition metal and A must be occupied by one of the metals like (Ca, Sr, Ba). Only Fe, Co, or Ni belong to B.[2-5]. Perovskite hydrides, denoted by ABH$_3$, have gained attention recently as a possible workable option for storing hydrogen. The following characteristics are typical of materials used to store hydrogen: These materials have a lot of hydrogen bonding. These materials have the following key characteristics: (1) they have a large amount of space that allows for the storage of a large amount of hydrogen; (2) they have catalytic properties that improve hydrogen absorption; and (3) Their hydrogen storage capability is enhanced by their respectable gravimetric hydrogen storage capabilities.. The gravimetric densities of hydroxide perovskites typically range from 1.2 to 6.0 weight percent. Metal hydride perovskites are a safer and more efficient way to store hydrogen than the liquid or compressed gas phase of hydrogen[6-9]. The study explores the different physical properties of perovskite hydrides, specifically KXH$_3$, with (X = Zn, Co) for hydrogen storage applications, using DFT approach for safer and more efficient storage.

## 2. Computational Details

The KXH$_3$ structure is cubic in nature and it is considered in all compositions using space group 221, assigning atomic positions as (0.5, 0.5, 0.5) for X, (0, 0, 0) for K and that of hydrogen (H$_1$, H$_2$ and H$_3$) are (0, 0.5, 0.5), (0.5, 0, 0.5) and (0.5, 0.5, 0) respectively. The structural properties of perovskites were investigated using Birch-Murnaghan's equation of state, adjusting energy vs. volume to identify the stable phase. These calculations employ 1000 k-points, an RMT of 7.0, and an energy cutoff of -6.0 Ryd. The physical properties of cubic compounds KXH$_3$ (X = Zn, Co) were calculated using the WIEN2k code after geometry optimization.[10]

## 3. Results and Discussion

This section will address the computation of various physical properties and hydrogen storage capabilities of potassium-based $KXH_3$ (X = Zn, Co) hydrides perovskites.

## 3.1 Structural Predictions

$KXH_3$ hydrides with (X = Zn, Co) are members of space group 221 and features a cubic crytal structure. The Wyckoff positions of K are (0, 0, 0), of (X = Zn, Co) are (0.5, 0.5, 0.5 ) and those of hydrogen (H1, H2, and H3) are (0, 0.5, 0.5), (0.5, 0, 0.5), and (0.5, 0.5, 0), respectively. The $KXH_3$ (X = Zn, Co) hydrides relaxed cubic crystal structure is shown in Figure 1. The energy versus volume plots in Figure 2(a–b) show these compounds minimum ground state energies. The B-E equation of states was used to investigate all physical properties of these compositions that are listed in Table 1.

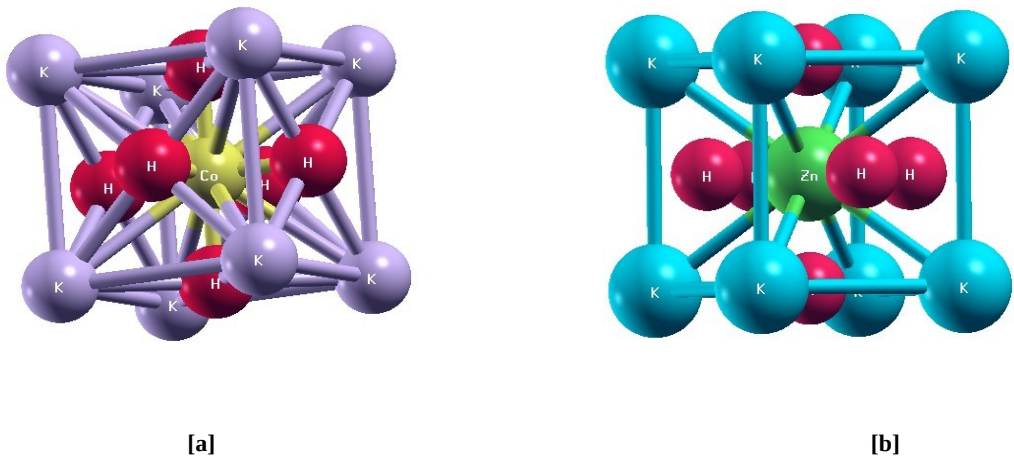

[a] [b]

Fig. 1. Relaxed optimised structures of [a] $KCoH_3$ [b] $KZnH_3$

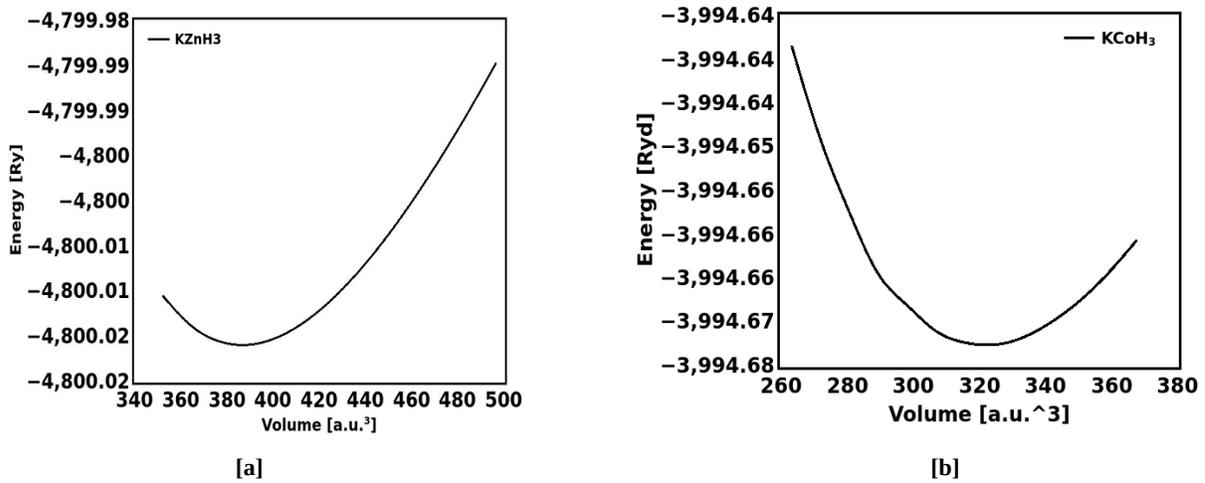

[a] [b]

Fig. 2. Energy vs volume graphs for [a] $KZnH_3$ [b] $KcoH_3$

Table1. Optimized structural and other parameters of $KXH_3$ hydrides with (X = Zn, Co)

| Perovskites | a (Å) | B | $V_0$ | B' | $E_0$ | $\Delta H_f$ | $T_d$ | $C_{wt\%}$ | t | μ |
|---|---|---|---|---|---|---|---|---|---|---|
| $KZnH_3$ | 3.8458 | 44.73 | 386.54 | 4.01 | -4800.02 | -198.72 | 1528.6 | 2.82 | 1.0 | .45 |
| $KCoH_3$ | 3.6250 | 67.32 | 321.45 | 4.21 | -3994.67 | -287.27 | 2209.8 | 3.00 | 1.0 | .45 |

## 3.2 Gravimetric storage capabilities, enthalpy, decomposition temperature, and stability of intermetallic hydrides

The decomposition temperature ($T_d$), enthalpy of formation ($\Delta h_f$) and the stability of intermetallic hydrides will be predicted using DFT methods[3]. Phase stability and dehydrogenation properties are found for $KXH_3$ (X = Zn, Co) hydrides perovskites. These values are calculated as following. The mechanisms that result in $KXH_3$ (X = Zn, Co) perovskites may be comprehended using the following equations as given below.

$$KZnH_3(s) \leftrightarrow KZn(s) + 3/2 * H_2(g) \quad \text{-------------- (1)}$$
$$\text{and}$$
$$KCoH_3(s) \leftrightarrow KCo(s) + 3/2 * H_2(g) \quad \text{-------------- (2)}$$

Using Hess's law as given below

$$\Delta H = \Sigma E_{tot}(produts) - \Sigma E_{tot}(reactants) \quad \text{-------------- (3)}$$

Our compositions' enthalpy of formation can be computed as

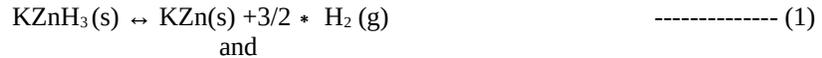
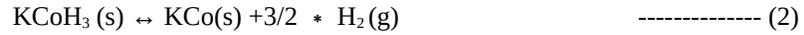

$$\Delta H_f = E_{tot}(KZnH_3) - E_{tot}(KZn) - 3/2 * E_{tot}(H_2) \quad \text{-------------- (4)}$$
$$\text{and}$$
$$\Delta H_f = E_{tot}(KCoH_3) - E_{tot}(KCd) - 3/2 * E_{tot}(H_2) \quad \text{-------------- (5)}$$

Employing the FP-LAPW approach we got formation of enthalpy of -287.27 kJ/mol.$H_2$ and -198.72 kJ/mol.$H_2$ for these hydrides. The exceptional thermodynamic stability of $KXH_3$ perovskites (where X=Zn, Co) is demonstrated by these values, which are significantly higher than the optimum value of -40 kJ/mol.$H_2$ according to US Department of Energy. Additionally, the traditional Gibb's energy ($\Delta G$) relation is expressed as follows:

$$\Delta G = \Delta H_f - T \Delta S \quad \text{-------------- (6)}$$

where $\Delta S$ represents entropy change, and $\Delta H_f$ is the formation enthalpy. This reltion is employed to determine the decompoistion temperature of $KXH_3$ (X = Zn, Co) hydrides perovskites. At equilibrium, the standards value of Gibb's energy is zero. Furthermore, the evolution of the hydrogen molecule during the dehydrogenation reaction is responsible for the value of the change in entropy, or $\Delta S$, which is given as $\Delta S = 130$ J/mol.$K^{-1}$. This relationship can be used to calculate the decomposition temperature given as .

$$T = \Delta H_f / 130. \quad \text{-------------- (7)}$$

The better stability of these compositions was confirmed when we found that the decomposition temperature for KZrH3 was 1529 K and for KCoH3 was 2209 K using the enthalpy of formation value from Table 1 in the aforementioned equation. The gravimetric percentage of the investigated compositions is crucial for their usefulness in investigating different hydrides for storage hydrogen and corresponding applications and is calculated using the following relation given as .

$$C_{wt\%} = [n_H * m_H / M_{(KXH3)} * 100] \% \quad \text{-------------- (8)}$$

Here, $m_H$, $n_H$ and $M(KXH_3)$ are the molar mass of a hydrogen atom, number of hydrogen atoms, and the molar mass of the host compound. Using the above equation we find the hydrogen storage percentage for these compositions in terms of gravimetric percentage as 2.82% and 3.00% as $KXH_3$ (X=Zn,Co). We have also investigated the value of the Goldschmidt factor (t) and the octahedral factor (μ) in order to determine the structural stability of these combinations. They are given by the following equations as mentioned below

$$t = r_K + r_H / 1.41[r_{(X = Zn, Co)} + r_H] \quad \text{-------------- (9)}$$

and

$$\mu = r_X / r_H \quad \text{-------------- (10)}$$

In these equations $r_k$ and $r_H$ are the radii of Potassium and Hydrogen while $r_X$ is the radii of (X = Zn, Co) respectively. The general value of t and µ for cubic perovskites structure lies in the range of 0.7-1 and 0.4 -0.7 respectively. Table 1 indicates that both $KXH_3$ (X = Zn, Co) have a value of t = 1.0 and a value of µ = 0.45, indicating that both compositions are structurally stable. when comparing the computed value of enthaply of formation ($\Delta H_f$) to the standard value of $\Delta H_f$, which is 40 kJ/mol.$H_2$, as suggested by the U.S. Department of Energy.

## **3.3 Electronic predictions.**

Perovskites, a crystal structure with exceptional electronic properties, are used in various applications like photodetectors, solar cells, LEDs, and hydrogen storage. Their tuneable bandgap allows for versatile applications[11-12]. Figures 3(a-b) show the energy band diagram for $KXH_3$ hydrides perovskites which are computed along initial Brillouin zone high symmetry points. The absence of an energy gap at the Fermi level (E.F.) between the conduction and valance bands suggests that the two compounds under investigation have metallic behaviour. T.D.O.S. and P.D.O.S. of $KXH_3$ have been calculated to better understand the electronic contribution, an essential idea in solid-state physics that represents the electronic state density at a given energy.

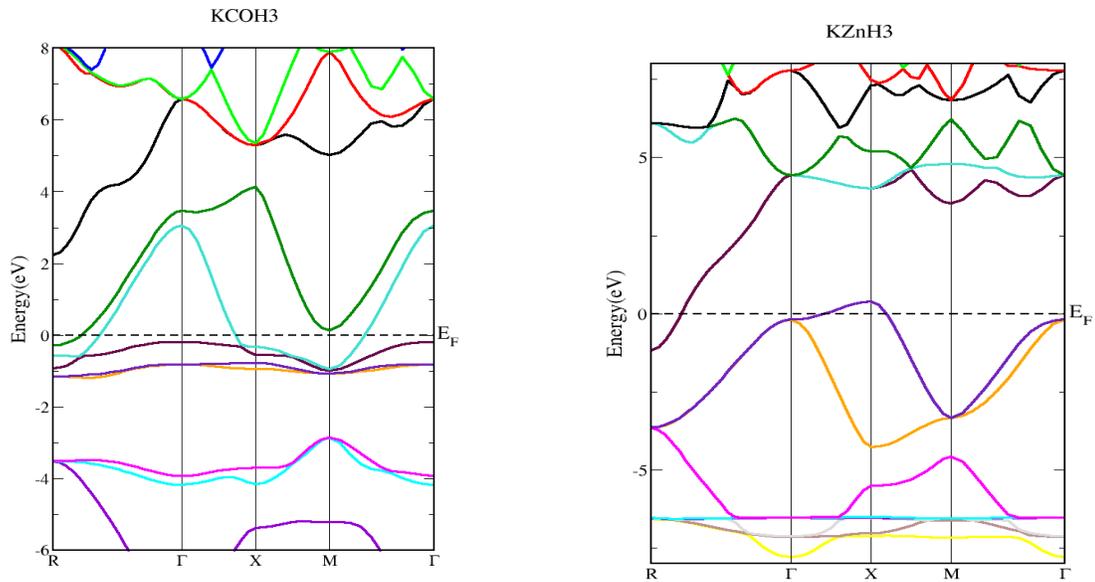

**Fig. 3. Energy Band structures for [a] KCoH3 [b] KZnH$_3$**

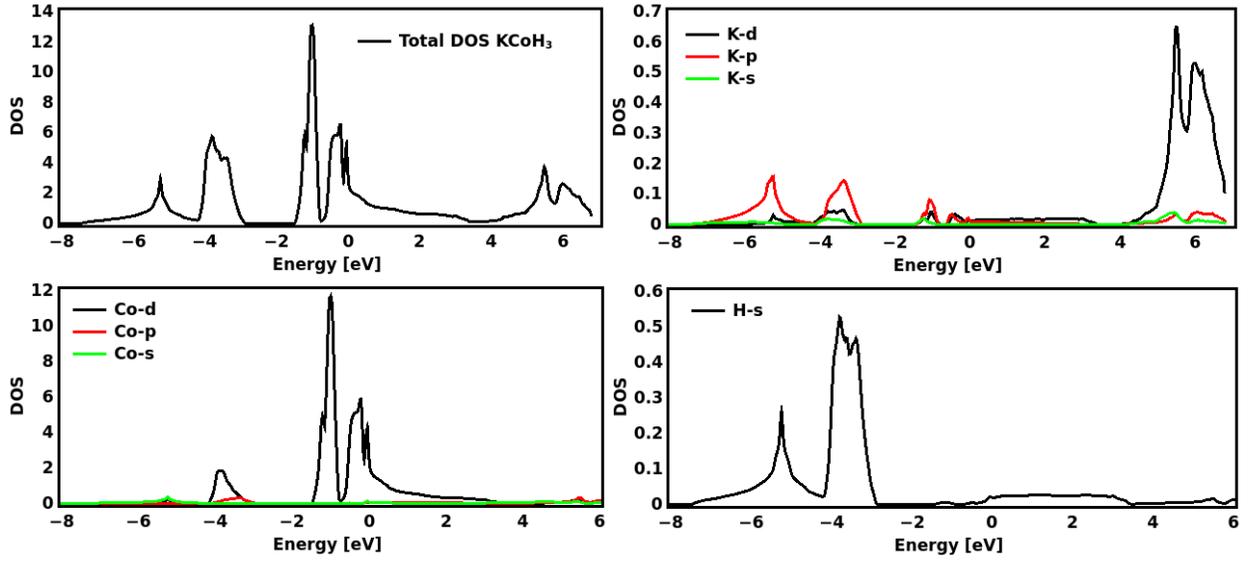

[a]

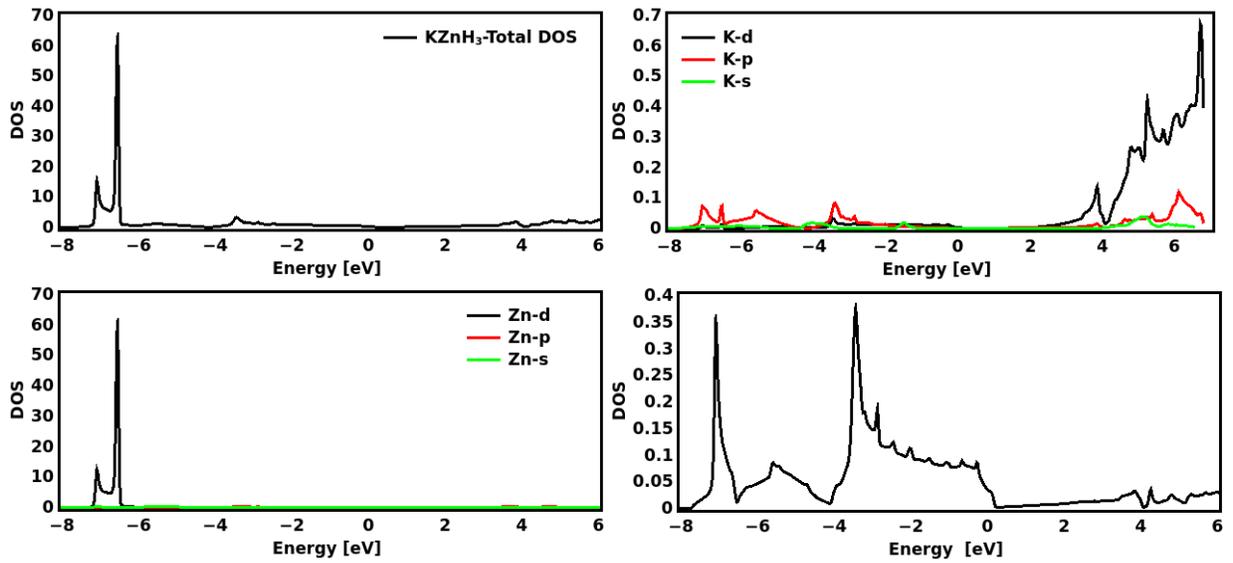

[b]

**Fig. 4. DOS/PDOS for [a] KCoH3 [b] KZnH$_3$**

Plots of T.D.O.S. and P.D.O.S. for KXH$_3$ (X = Zn, Co) hydrides perovskites are displayed in Fig. 4(a–b). The maximum T.D.O.S. values in the valence band for KZnH3 and KCoH3 are 60.5 states/eV and 12.5 states/eV, respectively. P.D.O.S., which analyzes the density of states in the orbital of a particular element or substance, is an essential instrument for comprehending the electrical structure of materials. The d-states of $KCoH_3$ and $KZnH_3$ significantly contribute to the band structure, with KCoH3's s- and p-states making significant contributions between -8 and -2 eV. The d-states of both compounds have made the largest contribution to the conduction band (0 to 6 eV).

## 4. Conclusion

Using the DFT-based WIEN2k code, the structural, electronic, stability, enthalpy, decomposition temperature and hydrogen storage of $KXH_3$ (X = Zn, Co) hydrides perovskites are investigated. According to the structural investigations, all compositions in space group Pm3m [no. 221] have a stable cubic structure. Plots of the density of states and energy band gap indicate that they are metallic in nature. The FP-LAPW approach indicates that $KXH_3$ (X = Zn, Co) perovskites have a thermodynamically stable enthalpy of formation of -198.72 kJ/mol.$H_2$ and 287.27kJ/mol.$H_2$, significantly higher than the optimal value of -40 kJ/mol.$H_2$. The investigated compositions show enhanced hydrogen storage cyclability and promising potential for hydrogen storage with gravimetric capacities of 2.82% and 3.0%.